\documentclass[aip,cha,reprint,noshowpacs]{revtex4-1}
\usepackage{graphicx}
\usepackage{hyperref}
\usepackage{amsmath,amssymb}
\usepackage{epstopdf}

\usepackage{amsmath}
\usepackage[usenames,dvipsnames]{color}

\graphicspath{{./figures/}{./new-figures/}}

\def\N#1{#1}
\def\Q#1{#1}
\DeclareMathOperator{\cotan}{cotan}

\begin{document}

\title{Continuous and discontinuous transitions to synchronization}
%\date{v9, \today, {changes are in red color}}
\author{Chaoqing Wang}
\affiliation{Department of Physics, East China Normal University, Shanghai, 200062, China}
\author{Nicolas B. Garnier}
%\affiliation{Laboratoire de Physique, CNRS UMR 5672, ENS de Lyon, Universit\'e de Lyon, France}
\affiliation{Univ Lyon, Ens de Lyon, Univ Claude Bernard, CNRS, Laboratoire de Physique, F-69342 Lyon, France}
\email{nicolas.garnier@ens-lyon.fr} 

\begin{abstract}
We describe how the transition to synchronization in a system of globally coupled Stuart-Landau oscillators changes from continuous to discontinuous when the nature of the coupling is moved from diffusive to reactive. 
We explain this drastic qualitative change as resulting from the co-existence of a particular synchronized macrostate together with the trivial incoherent macrostate, in a range of parameter values for which the later is linearly stable.
In contrast to the paradigmatic Kuramoto model, this particular state 
observed at the synchronization transition
contains a finite, non-vanishing number of synchronized oscillators, which results in a discontinuous transition.
We consider successively two situations where either a fully synchronized state or a partially synchronized state
exist at the transition.
Thermodynamic limit and finite size effects are briefly discussed, as well as connections with recently observed discontinuous transitions.
\end{abstract}

\pacs{05.45.Xt, 05.45.-a}

\maketitle

%% Leading paragraph

	When one considers coupled oscillators that are described not only with their phase, but also with their amplitude,
	the transition to synchronization is much richer and can be discontinuous. We show that this can be achieved by simply changing the nature of the coupling coefficient from dissipative to dispersive. We relate this phenomena to the co-existence of particular macroscopic solutions, which we qualitatively illustrate.

\section*{Introduction}

Synchronization of oscillators is a challenging topic which has received a lot of developments since the pioneering work of Kuramoto~\cite{Kuramoto:84}. Within this paradigm, an oscillator is described by its phase only; this is sufficient to describe the transition to synchronization of almost any ensemble of coupled oscillators in the limit of weak coupling~\cite{Kuramoto:review}.
If one wants to take into account other variables to describe more accurately the state of the oscillators, the most natural generalization of Kuramoto's paradigm is to consider not only the phase, but also the amplitude of the oscillators. The simplest model of such an oscillator is given by the Stuart-Landau equation, which is nothing but the normal form of a Hopf bifurcation~\cite{Cross:1993}.

Very few studies have been devoted to this much richer and generic case.
Despite detailed explorations~\cite{Strogatz:1991,Cross:2004,Cross:2006,Ott:2013}, coupled amplitude equations suffer from the inherent complexity associated with the doubling of the number of freedom degrees compared to Kuramoto phase equations. 
Of particular interest to us is the discontinuous ---~or first order~--- transition to synchronization~\cite{Cross:2004,Cross:2006,Strogatz:1991} that can be observed in this system.
% under very generic conditions. 
This feature is of great interest in complex systems to describe an abrupt change of their macroscopic behavior when only a single parameter is slightly changed.
%which may be associated with unpredictability.
%
Such a discontinuous transition has been documented in details in the Kuramoto-Sakaguchi phase model~\cite{Sakaguchi:86} under specific conditions on the randomness~\cite{Wolfrum:2012,Wolfrum:2013}, \N{as well as in the Kuramoto model with inertia~\cite{Tanaka:97,Acebron:2000} under generic conditions.}
It has also been previously observed in a Stuart-Landau system in the limit of weak coupling~\cite{Montbrio:2011}, when phase only is considered \N{and the dynamics reduces to the Kuramoto model}.

We first present the model used in this paper, a set of coupled Stuart-Landau equations, and detail the observation of either a continuous or discontinuous transition to synchronization, depending on the imaginary part of the coupling parameter. We then perform a complete linear stability of the incoherent macrostate to determine the position of transition point. We then further search for the existence of particular synchronized macrostate, in order to illustrate a scenario for the discontinuous transition.

We consider an ensemble of $N$ oscillators $\{z_j\}$, $1\le j\le N$. Each oscillator is coupled equally to all of the others:
\begin{equation}
\dot{z}_{j} = ( \mu+i \omega_{j}   ) z_{j} - B (1 + ic_2) | z_{j} |^{2}
   z_{j} +k e^{i  \alpha} ( \langle z \rangle -z_{j} ) , 
\label{eq:CGL}
\end{equation}
where
\begin{equation}  
\langle z \rangle \equiv \frac{1}{N} \sum^{N}_{j=1} z_{j} \equiv R e^{i\Psi}
\label{eq:mean_field}
\end{equation}
is the mean field. 
This model differs from the paradigmatic complex Ginzburg-Landau equation~\cite{Cross:1993,Hakim:1992} not only because of the non-local coupling, but also because of the non-uniformity of the oscillators.
As in the Kuramoto model, natural frequencies $\{\omega_j\}$ of the $N$ oscillators are randomly distributed with a probability density $g(\omega)$ which we choose unimodal for the sake of generality.
Varying $\alpha$ from 0 to $\pi/2$ changes the nature of the coupling from diffusive (dissipative) to reactive (dispersive)~\cite{Cross:2004,Cross:2006} 
and we show below that it also changes the nature of the transition to synchronization.

\section*{Model and Numerical observations}

We set $B=1$ without loss of generality, while choosing $c_2=0$ for the sake of simplicity. So we consider a real coefficient for the non-linear term, and a complex coefficient for the coupling term, in contrast to \cite{Strogatz:1991,Cross:2004,Cross:2006}. In this case, the dynamics of oscillator $j$ obeys:
\begin{equation}
\dot{z}_{j} = ( \mu+i \omega_{j}   ) z_{j} - | z_{j} |^{2}
   z_{j} +k e^{i  \alpha} ( \langle z \rangle -z_{j} ) \,.
\label{eq:CGL.simple}
\end{equation}

This model system with $c_2=0$ has been derived exactly in the context of laser arrays~\cite{Jiang:1992}.
In the following, we use a Lorentzian distribution $g(w)=({\sigma}/{\pi})/{(\omega^{2} + \sigma^{2})}$ with $\sigma=1/\pi$ and unless noted otherwise $\mu=1$ and $N=1000$. Numerical simulations were also conducted with a Gaussian distribution without any qualitative changes to be noted.

We use the modulus $R \equiv |\langle z\rangle|$ of the mean field as an order parameter~\cite{Strogatz:1991,Cross:2006} and we consider the coupling strength $k$ as the control parameter. The width $\sigma$ of the frequency distribution and the angle $\alpha$ are other relevant parameters but we keep $\sigma$ fixed in this study and focus first on two typical values of $\alpha$ : a smaller one (0.5) and a larger one (1.3). For small coupling $k$, the system is expected to be in an incoherent macrostate, while in the limit of large coupling strength $k$, it is expect to be in a synchronized state. We restrict ourselves to $k<\mu/\cos\alpha$.

\begin{figure}[tb]
\begin{center}
\includegraphics[width=\linewidth]{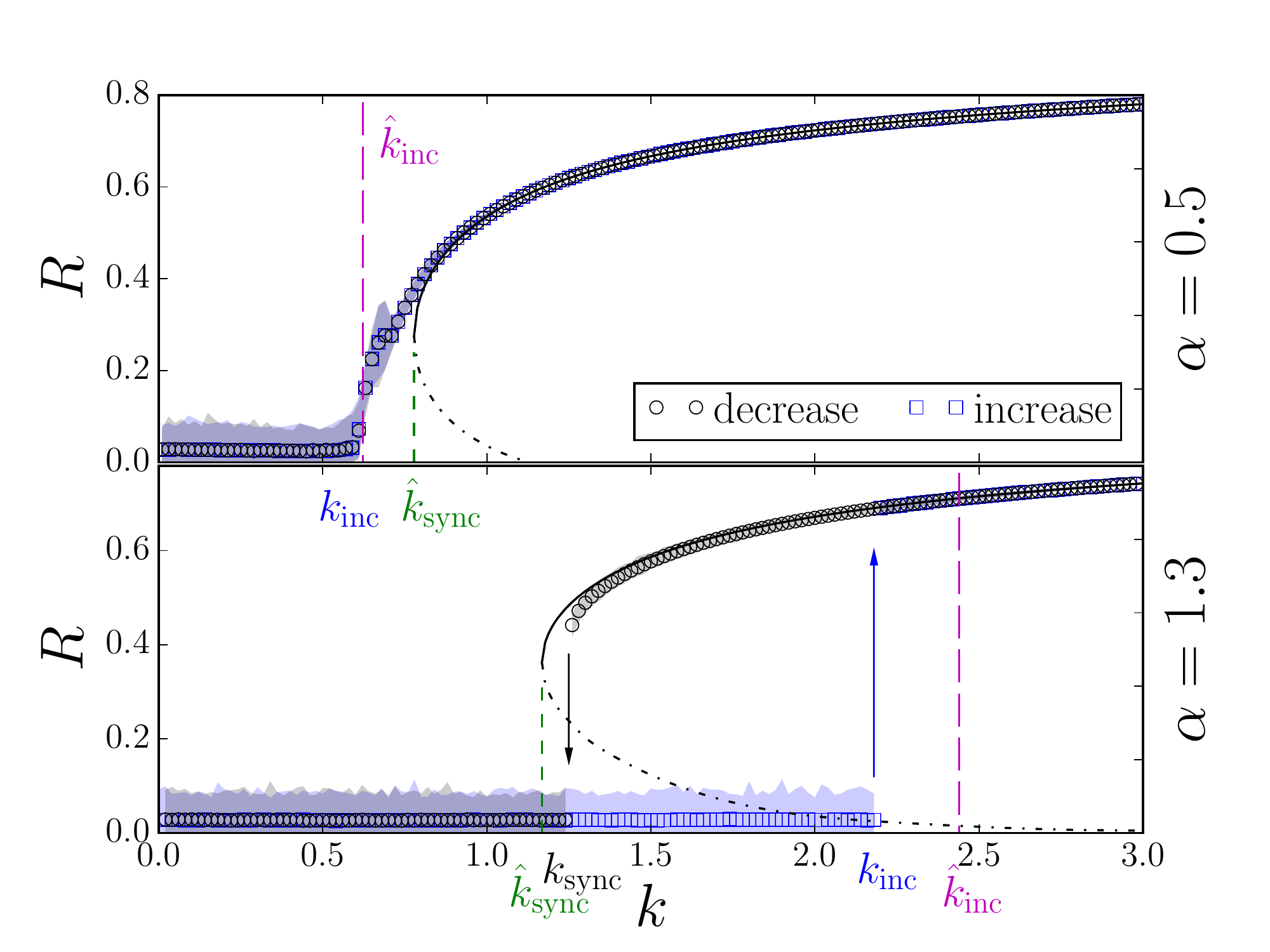}
\end{center}
\caption{Bifurcation diagram for two typical values of $\alpha$, showing how the order parameter $R$ varies with the control parameter $k$. 
Symbols indicates DNS values of the time average of ${R}$, while the range of time fluctuations is represented by color shades. Blue squares and blue shade, resp. black circles and grey shade, correspond to macrostates obtained by adiabatically increasing, resp. decreasing, the coupling $k$.
Upper panel for small $\alpha=0.5$ shows a continuous transition at the critical value $k_{\rm inc}\approx0.58$.
Lower panel for large $\alpha=1.3$ shows hysteretic behavior between $k_{\rm sync}\approx1.25$ and $k_{\rm inc}\approx2.18$ .
Magenta, resp. green, vertical dashed lines indicates the estimates $\hat{k}_{\rm inc}$, resp. $\hat{k}_{\rm sync}$.
Continuous curves (plain and dashed) show the fully synchronized solutions of the self-consistency eqs.(\ref{eq:autocoherence:real},\ref{eq:autocoherence:imag}). 
\label{fig:bifurc}
}
\end{figure}

We perform direct numerical simulations (DNS) of eqs.(\ref{eq:CGL},\ref{eq:mean_field}), using a fourth-order Runge-Kutta scheme with time step 0.01. Transient states are discarded by waiting 20 000 time units and we average the order parameter $R$ over 20 000 time units.
The dependence of the time-averaged order parameter on the coupling $k$ is presented in Fig.~\ref{fig:bifurc} for two typical values of $\alpha$. 
For smaller values of $\alpha$ (diffusive coupling), a single state is observed whatever the initial conditions are and the bifurcation is continuous. 
On the contrary, for larger values of $\alpha$ (reactive coupling), there exists a range of coupling $k$ where at least two distinct macrostates are observed, depending on the initial conditions. In order to enlarge this multistability range as much as possible, we use the following protocol in the DNS: starting with a low value of $k$ where only the incoherent state is observed, we increase adiabatically the coupling by $\delta k=0.01$ and use the last microstate of the system at $k$ as the initial conditions for the simulation at $k+\delta k$. Conversely, starting from large values of the coupling where only a synchronized state is observed, we adiabatically reduce the coupling to $k-\delta k$ while using the last microstate at $k$. This allows us to track the hysteresis region associated with the discontinuous transition.

For any value of $\alpha$, we define $k_{\rm inc}$ as the largest value of the coupling $k$ for which the completely incoherent solution is observed. Above $k_{\rm inc}$, a synchronized solution is observed and the order parameter is strictly positive. 
For larger $\alpha$, we additionally define $k_{\rm sync}$ as the smallest value of $k$ for which the synchronized solution is observed in the hysteretic region. Adiabatically increasing $k$ leads to a transition from the incoherent state at $k_{\rm inc}$, and adiabatically reducing $k$ leads to a transition from the synchronized state either at $k_{\rm inc}$ if the transition is continuous, or at $k_{\rm sync} < k_{\rm inc}$ if the transition is discontinuous.

\medskip

\begin{figure}[tb]
\begin{center}
\includegraphics[width=\linewidth]{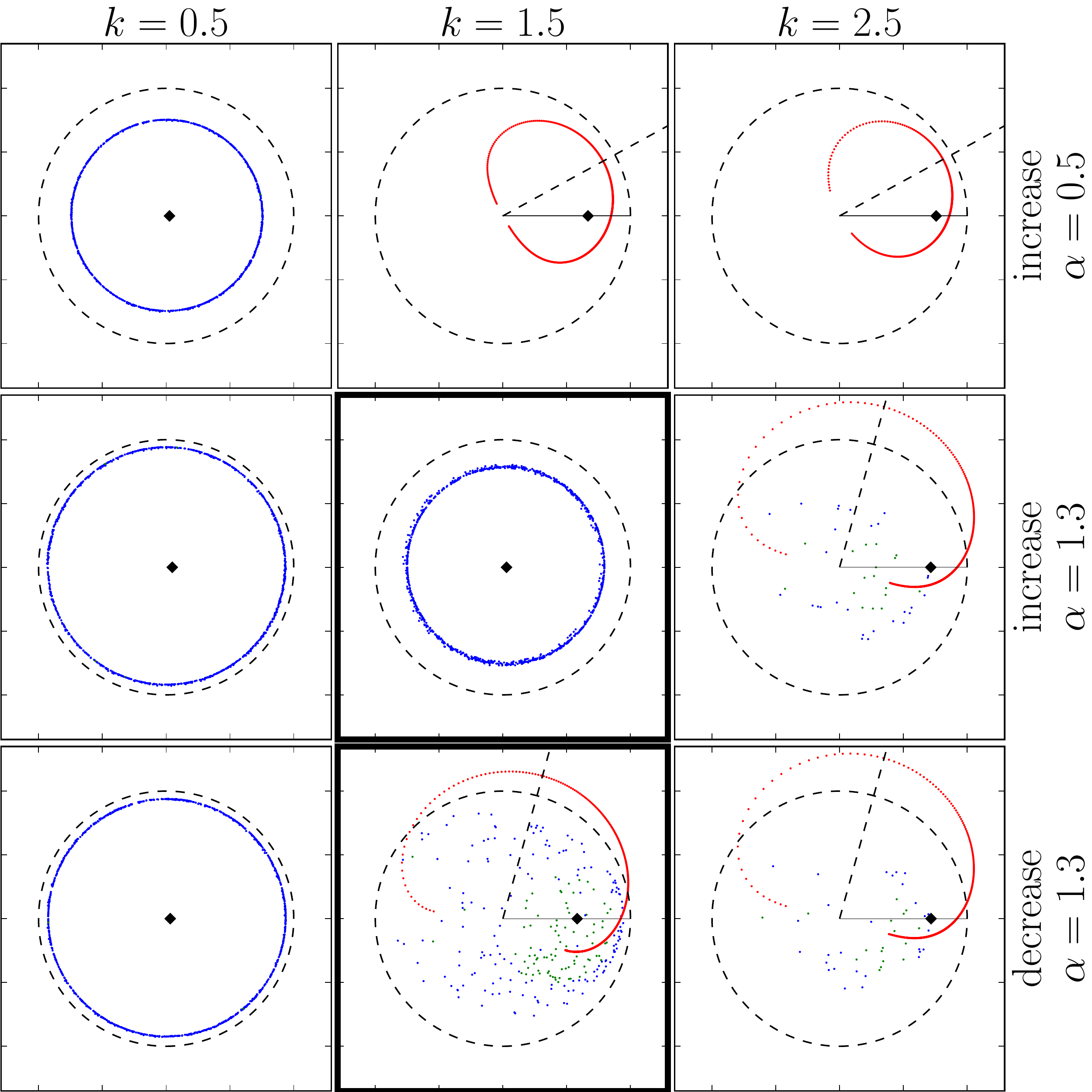}
\end{center}
\caption{Snapshots --- instantaneous microstates --- in the complex plane for typical $\alpha, k$. 
\N{Blue, red and green dots represent the incoherent, fully synchronized (locked) and modulated oscillators, respectively.}
First line: for smaller $\alpha$ (0.5 here) no hysteresis is observed. 
Second and third lines: for larger $\alpha$ (1.3 here), multistability is observed.
The three typical coupling strengths $k$ were chosen such that $0.5<k_{\rm sync}<1.5<k_{\rm inc}<2.5$.
The thick bordered sub-panels show two different states observed inside the hysteresis depending on initial conditions. 
Black dashed circles are unit circles (radius 1).
All microstates have been rotated so that $\Psi=0$, and the mean-field is represented by a black diamond on the real axis, while the angle $\alpha$ is indicated by a dashed line.
%
%As for all figures in this article, blue, resp. black, color indicates data obtained by adiabatically increasing, resp. decreasing, the coupling.
}
\label{fig:snapshots}
\end{figure}

Microstates corresponding to typical macrostates are presented in Fig.~\ref{fig:snapshots}. 
In the incoherent macrostate, $\langle z\rangle=0$ and all oscillators are independent. They all rotate at their own \N{individual} angular velocity \N{$\omega_j-k\sin\alpha$ ---~their natural frequency shifted by $k\sin\alpha$~---} on the \N{same} circle of radius $r_0\equiv\sqrt{\mu-k\cos\alpha}$. 
It is worth mentioning here that both $k_{\rm inc}$ and $k_{\rm sync}$ are always smaller than $\mu/\cos\alpha$, so $r_0$ is always defined in the range of $k$ that we study.
In synchronized macrostates with large $R$, a large fraction of oscillators are located in a cluster which rotates with the angular velocity $\Omega \equiv \partial \Psi/\partial t$. In the referential of the mean field, obtained by multiplying all oscillators by $e^{-i\Psi}$, the envelope of this cluster is symmetrical around the line defined by the polar angle $\alpha$. Besides oscillators lying on the cluster, unsynchronized oscillators are observed. The number of such oscillators increases when $\alpha$ is increased or $k$ is reduced. In the bistable regime, for some fixed set of parameters $(\alpha, k)$, the macrostate can be either incoherent or synchronized, depending on initial conditions. 

To document how the transition can be continuous or discontinuous, we study successively the linear stability of the incoherent state and the existence of either a fully synchronized state or a partially synchronized state with a majority of synchronized oscillators.

\section*{Linear stability of the incoherent macrostate}

When the coupling $k$ is increased from 0, the observed macrostate is incoherent\N{, see Fig.\ref{fig:snapshots}, first column}. Each oscillator rotates freely on the circle of radius $r_0$ and $\langle z \rangle =0$, up to statistical fluctuations~\cite{Kuramoto:84,Cross:2006} of order $N^{-1/2}$. 
The linear stability of this specific macrostate is conveniently studied in the thermodynamic limit $N\rightarrow\infty$ using the Ott-Antonsen's method~\cite{Ott.Antonsen:2008,Ott:2013} 
to get an estimate $k_{\rm inc}^\infty$ of the critical value  for which this incoherent state loses stability. 

Dropping the index $j$ and replacing $z_j(t)$ by $z(\omega,t)=r e^{i\theta}$, eq.~(\ref{eq:CGL.simple}) is rewritten in polar coordinates:
\begin{eqnarray} 
\dot{r} & = & (r_0^2-r^{2})r+\frac{kR}{2}(e^{i(\Psi-\theta+\alpha)} + e^{i (\theta-\Psi-\alpha)})  \,, \label{eq:CGL.radius} \\
\dot{\theta} & = & \bar{\omega} +\frac{kR}{2}(e^{i(\Psi-\theta+\alpha)} - e^{i (\theta-\Psi-\alpha)})   \,, \label{eq:CGL.phase}
\end{eqnarray}
where $\bar{\omega}=\omega-k \sin \alpha$ is the oscillator natural frequency shifted by the reactive part of the coupling.
A macrostate is described by $f(\omega,\theta,r,t)$, a time dependent joint density of the state variables $(r,\theta)$ and parameter $\omega$.
The incoherent macrostate is time independent and because $R=\langle z \rangle_0 = 0$, the radius $r$ of any oscillator equals $r_0$; the corresponding density $f_0$ can be written as~\cite{Ott:2013}
\begin{equation*}
f_{0} ( \omega, \theta ,r ) = \frac{g ( \omega )}{2 \pi} \delta \left( r- r_0 \right) \,.
\label{eq:unsynchronized.state.density}
\end{equation*}
To probe the linear stability of the incoherent macrostate, the density $f_0$ is perturbed by writing $f(\omega, \theta ,r,t) = f_0 + f_1(\omega,r) e^{-i\theta+st}$ while noting that non-resonant terms can be discarded~\cite{Ott:2013}.
This leads to a perturbed mean field that we write $\langle z \rangle = \langle z \rangle_0 +\langle z \rangle_1 e^{st}$ where $\langle z \rangle_1$ is the amplitude of the linear perturbation:
$$
\langle z \rangle_1 = \int^{+ \infty}_{- \infty}d\omega \int^{+\infty}_{0}dr \int_{0}^{2\pi}d\theta  
\left( z f_1 ( \omega,r ) e^{-i\theta} \right) \,.
% {+st}
$$ 

The conservation of density in phase space requires that $f$ satisfies the continuity equation
\begin{equation*}
\frac{\partial f}{\partial t}+ \frac{\partial}{\partial r} (f\dot{r}) + \frac{\partial}{\partial \theta} (f\dot{\theta})=0 \,,
\label{eq:continuity-equation}
\end{equation*}
from which we deduce using eqs.(\ref{eq:CGL.radius},\ref{eq:CGL.phase}) that
\begin{multline}
(s -i \bar{\omega}) f_1 + \frac{\partial}{\partial r} [ ( r_0^2 -r^{2})r f_1  ] = \\
\frac{1}{4\pi} k e^{i \alpha} \langle z \rangle_1  g ( \omega )
\left[ \frac{\delta ( r- r_0)}{r_0} - \frac{d}{d r}  \delta ( r- r_0 ) \right] \,,
\label{eq:continuity:linear}
\end{multline}

We then use the antsatz from ~\cite{Ott:2013} to express the amplitude of the perturbation $f_1$ as both proportional to $g(\omega)$ and as a linear combination of $\delta \left( r- r_0 \right)$ and its first derivative in $r$:
\begin{equation*}
%\label{eq:antsatz}
f_1 \equiv \frac{g(\omega)}{4\pi}k \langle z \rangle_1  e^{i \alpha} [c_1 \delta (r-r_0) + c_2 {\frac{d}{dr}  \delta (r-r_0)}]
\end{equation*}
We solve eq.(\ref{eq:continuity:linear}) in $(c_1,c_2)$ to obtain:
\begin{equation*}
f_1= \frac{1}{4\pi} k e^{i \alpha} \langle z \rangle_1 g (\omega) 
\left[ \frac{\delta ( r- r_0)}{( s -i \bar{\omega}) r_0}  -\frac{\frac{d}{d r}  \delta ( r- r_0 )}{s -i \bar{\omega}+2 {r_0^2}} \right]
\end{equation*}

We then inject this expression of the perturbed density in the equation defining the perturbed mean field $\langle z \rangle_1$. We obtain a self-consistency equation which is linear in the perturbation and which reduces to the following equation for the growth rate $s$:
\[ \frac{1}{k e^{i \alpha}} = \int^{+\infty}_{-\infty} g(\omega)  
\frac{s+{r_0}^2-i\bar{\omega}}{( s-i\bar{\omega}) ( s+2r_0^2-i\bar{\omega})} d \omega
\]
.%where $\bar{\omega}=\omega-k\sin\alpha$ is the oscillator natural frequency shifted by the reactive part of the coupling.

\begin{figure}[tb]
\begin{center}
\includegraphics[width=\linewidth]{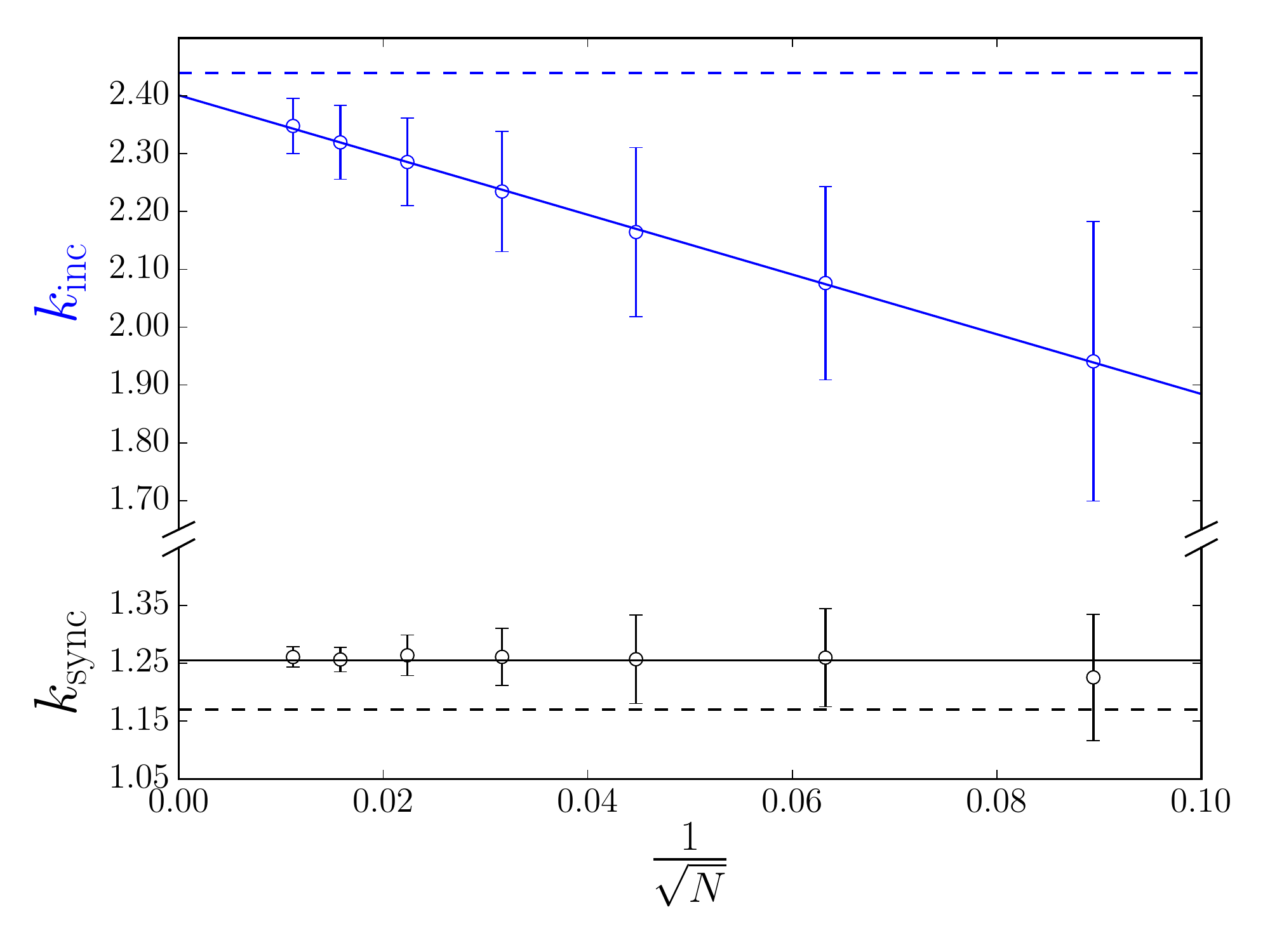}
\end{center}
\caption{Dependence of $k_{\rm inc}$ and $k_{\rm sync}$ on the system size $N$, for $\alpha=1.3$.
For each value of $N$, 100 different realizations of the frequencies $\{\omega_j\}_{1\le j\le N}$ have been used. 
Circles, resp. error bars, represent the average, resp. standard deviation, over the realizations. The continuous blue line is a linear fit, and the continuous black line represents a constant.
The blue dashed line is the value $\hat{k}_{\rm inc}$ from eq.(\ref{eq:kc1}) for $N=\infty$, and the black dotted line is the estimate $\hat{k}_{\rm sync}$ obtained by solving eqs.(\ref{eq:autocoherence:real},\ref{eq:autocoherence:imag}).}
\label{fig.size_effect}
\end{figure}

For a Lorentzian frequency distribution, the integral can be computed explicitly using residues theorem. We then track the onset of instability by setting the real part of the growth rate $s$ to zero. Eliminating the imaginary part of $s$ between the real and imaginary part of the complex equation then gives the following relation between parameters at onset:
\begin{equation}
\left[(\tan\alpha)^2 \frac{(2k'-1)k'^2}{(2-3k')^2} + k'-2\sigma'\right](1-k') + \sigma'^2 = 0
\label{eq:quartic.equation.for.lorentzian}
\end{equation}
where $k'= {k \cos\alpha}/({\mu + \sigma})$ and $\sigma'={\sigma}/({\mu+\sigma})$. 
In the limit of large $\alpha \rightarrow \pi/2$, this quartic equation in $k'$ has the four asymptotic solutions $\{0, 0, 1/2, 1\}$. The interesting one is the smallest non-vanishing one (1/2), and an expansion in $\cotan\alpha$ leads to the approximate expression:
\begin{equation}
 \hat{k}_{\rm inc} \cos\alpha = \frac{\mu+\sigma}{2} - \frac{(\mu-\sigma)^2}{4(2\sigma + (\mu+\sigma)(\tan\alpha)^2)}
 %+{\cal O}(1/\tan\alpha^2)
\label{eq:kc1}
\end{equation}
which gives $\hat{k}_{\rm inc}$ as a function of $\alpha$ and $\sigma$.
Although derived for $\alpha \rightarrow \pi/2$, eq.(\ref{eq:kc1}) gives a surprisingly good estimate for $\alpha=0$, where eq.(\ref{eq:quartic.equation.for.lorentzian}) becomes quadratic and can be solved exactly, leading to $\hat{k}_{\rm inc}=0.51$. Comparison of formula (\ref{eq:kc1}) with DNS for $\alpha=1.3$ is shown in Fig.\ref{fig.size_effect}. Agreement is good in the continuum limit, but eq.(\ref{eq:kc1}) overestimates the critical value; this may be due to our considering a specific form for the perturbation \N{$f_1$}.% as eq.(\ref{eq:antsatz}).
% \N{=  f_1(\omega,r) e^{-i\theta+st}}$, \N{containing only the leading Fourier mode in $\theta$; Incorporating higher orders terms of the Fourier expansion would result in a larger instability range, and a smaller, more accurate, estimate $\hat{k}_{\rm inc}$}. 
For smaller systems, the DNS value of $k_{\rm inc}$ not only diminishes, but it also fluctuates more intensely from one realization of the frozen disorder $\{\omega_j\}$ to another.

\section*{Existence of a fully synchronized macrostate}

\begin{figure}[tb]
\begin{center}
\includegraphics[width=\linewidth]{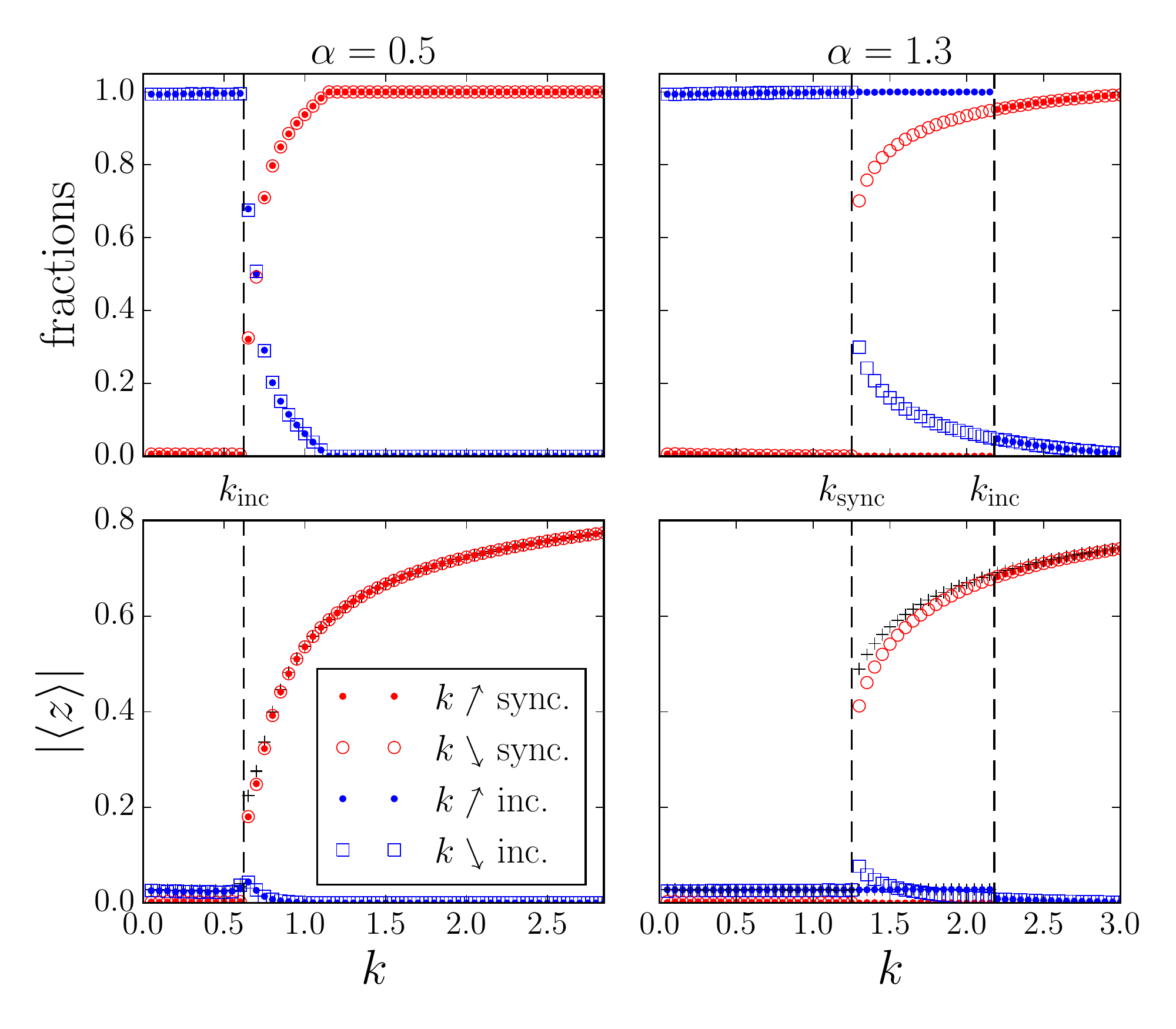}
\end{center}
\caption{
\N{
First row: fractions of synchronized oscillators ($\circ$, red) and incoherent oscillators ($\Box$, blue) as a function of the coupling $k$ in the DNS.
Second row: modulus of the contribution to the mean field computed from the population of synchronized oscillators (red) and from incoherent oscillators (blue). The total mean field (same data as in Fig.~\ref{fig:bifurc}) is 
represented by black symbols (+) for comparison. 
Open symbols represent a DNS with increasing $k$ and dots a decreasing $k$.
Left column is for $\alpha=0.5$ (continuous transition) and right column is for $\alpha=1.3$ (hysteresis between $k_{\rm sync}$ and $k_{\rm inc}$).}
}
\label{fig.fractions}
\end{figure}

When the coupling $k$ is decreased from arbitrarily large values, the order parameter $R(k)$ is finite and the observed macrostate seems to be coherent. Nevertheless, microstates are rather complex and individual oscillators can be in at least three different regimes: \N{fully synchronized (colored in red in Fig.\ref{fig:snapshots}), modulated (green in Fig.\ref{fig:snapshots}), or incoherent (blue in Fig.\ref{fig:snapshots})}. 
 Fully synchronized oscillators have a constant radius and they all rotate at the angular velocity $\Omega$ of the mean field, thus defining the synchronized cluster. Modulated oscillators have a time-averaged angular velocity equal to $\Omega$
 %~\cite{Wang:2015} 
 and we treat them in the following as fully synchronized, because they contribute to the mean field $\langle z \rangle$ exactly as if they were fully synchronized. Incoherent oscillators have an angular velocity which is far from, and incommensurate with, $\Omega$. 
\N{We have checked that they do not contribute significantly to $\langle z \rangle$. 
Fig.~\ref{fig.fractions} shows the dependence in the coupling $k$ of the fractions of synchronized oscillators ---~whether fully synchronized or modulated~--- and incoherent oscillators, together with their respective contribution to the mean field. When the transition is continuous ($\alpha=0.5$), although an arbitrary large fraction of oscillators are not synchronized around the transition at $k_{\rm inc}$, their contribution to the mean field is always negligible. The same conclusion holds around both threshold values $k_{\rm sync}$ and $k_{\rm inc}$ when the transition is discontinuous ($\alpha=1.3$).}
%
%%% new figure
Reducing $k$ increases the number of modulated oscillators
%~\cite{Wang:2015} 
and more importantly increases the number of incoherent oscillators, which as a result reduces $R$.

We now focus on a particular macrostate composed only of synchronized oscillators. We search for microscopic states $\{z_j\} = \{r_j e^{i\theta_j}\}$ and impose that $r_j$ and $\theta_j-\Psi$ are time independent.
For each oscillator, the two variables $(r_j, \theta_j%-\Psi
)$ are solutions of a set of two equations
\begin{eqnarray} 
%\dot{r}_j 
0& = & (r_0^2-r_j^{2})r_j+kR \cos(\Psi - \theta_j + \alpha )  \label{eq:polar:radius} \,, \\
%\dot{\theta}_j 
0& = & (\omega_j-\eta) +k \frac{R}{r_j} \sin(\Psi-\theta_j + \alpha ) \label{eq:polar:phase}  \,,
\end{eqnarray}
where $\eta=\Omega+k\sin\alpha$ is the frequency shift.
\N{For a given mean field $(R,\Omega)$, solving eqs.(9) and (10) gives $r_j(\omega)$ and $\theta_j^0(\omega) \equiv \theta_j(\omega, t) - \Omega t$, and hence the enveloppe $r(\theta)$ of the synchronized cluster (as seen in red in Fig.\ref{fig:snapshots}). }
We note here an interesting difference from the Kuramoto equation, where only the phase of oscillators is taken into account and the dynamics is described by eq.~(\ref{eq:polar:phase}) with a fixed radius $r_j=1$. Thanks to the radius variable $r_j$, an arbitrary large number of Stuart-Landau oscillators can be synchronized for arbitrarily small values of $R$: for any natural frequency $\omega_j$, a solution of (\ref{eq:polar:phase}) may exist, albeit with a small radius $r_j$ (see next section).

Eqs.(\ref{eq:polar:radius}) and (\ref{eq:polar:phase}) can be combined into a single equation for the radius only. Dropping the index $j$ and defining $\rho(R,\Omega,\omega)=r^2/b$ the normalized squared radius with $b=r_0^2 / 3$, $\omega'=(\omega-\eta)/b$ the normalized frequency and $R'^2=k^2R^2/4b^3$ the normalized order parameter, this equation reads
\begin{equation}
(\rho-2)^3 + p(\rho-2) + q = 0 \,,
\label{eq:cubic:rho}
\end{equation}
where $p=\omega'^2-3$ and $-q/2=2R'^2-1-\omega'^2$.
\N{The corresponding phase $\theta^0$ can then obtained by dividing eq.(\ref{eq:polar:phase}) by eq.(\ref{eq:polar:radius}):
$$ \tan(\alpha-\theta^0) = \frac{\omega'}{1-\rho} \,.
$$}

The definition (\ref{eq:mean_field}) of the mean field gives a complex self-consistency equation to be satisfied by the parameters $(R,\Omega)$ or their normalized expressions $(R',\eta)$. 
\N{In the thermodynamic limit, (\ref{eq:mean_field}) can be rewritten as:
$$
R e^{-i\alpha} = \int {\rm d}\omega g(\omega) r e^{i(\theta-\Psi-\alpha)} \,.
$$
Using eqs (\ref{eq:polar:radius},\ref{eq:polar:phase}), the complex exponential in the integral is then replaced by $\frac{r}{kR}(r^2 - r_0^2 + i(\omega-\eta))$ to get an expression which does not depend explicitly on the angle variable $\theta$. Using the normalized variables $\rho$ and $R'$, the real and imaginary parts of this expression read}:
\begin{eqnarray}
4R'^2\frac{b}{k} \cos\alpha
&=& {\int_{-\infty}^{+\infty} d\omega} g(\omega) \rho(\rho-3) \,,
\label{eq:autocoherence:real} \\
-4R'^2\frac{b}{k} \sin\alpha
&=& {\int_{-\infty}^{+\infty} d\omega} g(\omega) \rho \N{\omega'} \,.
\label{eq:autocoherence:imag}
\end{eqnarray}
As it is usual with amplitude equations, the imaginary part (\ref{eq:autocoherence:imag}) contains the information on the phase~\cite{Cross:2006} ---~here via the explicit dependence of $\omega'$ in the phase shift $\eta$~--- and it can formally be used to obtain  $\eta(R')$ or equivalently $\Omega(R)$. 
Eq.(\ref{eq:autocoherence:real}) with $\Omega(R)$ then determines implicitly the order parameter $R$.

A detailed examination of eq.(\ref{eq:cubic:rho}) shows that its discriminant $(q/2)^2+(p/3)^3$ is positive for all $\omega'$ as long as $R'>\sqrt{2}$. In that case, the cubic equation (\ref{eq:cubic:rho}) has a single real solution 
$\rho(\omega',R',\eta)$ that can be expressed analytically, for any oscillator in the distribution.
A fully synchronized state can therefore exist.
For large values of $\alpha$, when the transition is discontinuous, we always measure $R'>\sqrt{2}$ in the DNS.
We only observe $R'<\sqrt{2}$ when the transition is continuous.

To get a more precise insight, we numerically solve eqs.(\ref{eq:autocoherence:real},\ref{eq:autocoherence:imag}) in $(R,\Omega)$ after performing the integrals over the frequency distribution.
To do so, we express $\rho(\omega',R',\eta)$ using the real solution of (\ref{eq:cubic:rho}) that exists for all values of the discriminant. For negative values of the discriminant, this solution turns out to be the largest one among the three possible ones. As a consequence, we obtain the largest possible value of $R'$, according to (\ref{eq:autocoherence:real}).
Solutions are plotted as continuous and dashed lines in Fig.\ref{fig:bifurc}. We measure the value $\hat{k}_{\rm sync}$ of the coupling $k$ below which no solution of eqs(\ref{eq:autocoherence:real},\ref{eq:autocoherence:imag}) exist. 

\N{This method assumes the existence of a fully synchronized macrostate, so that the integrals (\ref{eq:autocoherence:real}) and (\ref{eq:autocoherence:imag}) can be evaluated over the all real axis.
As a consequence, it is assumed not only that the solution $\rho(\omega,R',\eta)$ exists (it does), but also that it is stable. This is unfortunately not the case close to the critical points $k_{\rm inc}$ or $k_{\rm sync}$, 
where the percentage of synchronized oscillators is always less than 100\% in the DNS (see Fig.\ref{fig.fractions}).
}
Nevertheless, this method gives a good estimate for the coupling parameter at which the synchronized solution disappears. Moreover, it does not depend on the system size, exactly as observed in the DNS.
Interestingly, we obtain a value $\hat{k}_{\rm sync}$ for any value of $\alpha$, including smaller ones for which the DNS reveals a continuous transition but in this case $\hat{k}_{\rm sync}>k_{\rm inc}$.
\N{This is in agreement with DNS, as we indeed observe fully synchronized states for $\alpha=0.5$ when $k$ is large enough: all the oscillators are then locked to the mean field (see first row of Fig.\ref{fig:snapshots} and Fig.~\ref{fig.fractions}).} 

\section*{Partially synchronized states}

Microstates corresponding to a synchronized macrostate can of course be composed of a fraction of synchronized oscillators and a fraction of incoherent oscillators. These partially synchronized ---~or mixed~--- states, considered in the original formulation of Kuramoto~\cite{Kuramoto:84}, are expected to be always present in a range of coupling coefficient, contrary to the fully synchronized macrostate which does not exist in the vicinity of the critical point in the Kuramoto model. 

When the transition is continuous, we observe in the DNS that the fraction of synchronized oscillators vanishes when $k$ approaches $k_{\rm inc}$ from above \N{(see Fig.~\ref{fig.fractions} for $\alpha=0.5$)}. The mixed states then contain a small fraction of synchronized oscillators, and a large fraction of non-synchronized oscillators, which makes the order parameter $R$ fluctuate strongly, as can be seen in Fig.\ref{fig:bifurc} above $k_{\rm inc}$ for $\alpha=0.5$.
When the transition is discontinuous, we observe in the DNS that when $k$ is reduced down to $k_{\rm sync}$ the \N{percentage} of synchronized oscillators in the synchronized state is reduced, but remains always larger than about 50\% \N{before jumping down to 0 (see Fig.~\ref{fig.fractions} for $\alpha=1.3$)}.

%This hints at deriving analytically a lower bound for the existence of mixed states with a majority of synchronized oscillators.}

When solving numerically the self-consistent relations eqs.(\ref{eq:autocoherence:real},\ref{eq:autocoherence:imag}), we made an additional observation: on the synchronized branch above $k_{\rm sync}$ the discriminant of the cubic equation (\ref{eq:cubic:rho}) is positive for all frequencies in the distribution. 
We therefore conjecture that synchronized oscillators have a positive discriminant, and as a consequence a single solution to their cubic equation.
We checked in the DNS that the synchronized macrostates for larger values of  $\alpha$ are indeed composed only of oscillators which have a positive discriminant in their eq.(\ref{eq:cubic:rho}). 
When $\alpha$ is small and the transition is continuous, we observe in the DNS in a range of values of $k$, between 
$k_{\rm inc}$ and $\hat{k}_{\rm sync}$ that the number of synchronized oscillators is growing continuously from 0, as does $R'$. Interestingly, above a value close to $\hat{k}_{\rm sync}$, we have $R'>\sqrt{2}$ in the DNS and all oscillators have a positive discriminant. This motivates the following counting of oscillators.

Looking at eq.(\ref{eq:cubic:rho}), we note that if $p>0$, {\em i.e.}, $|\omega'| > \sqrt{3}$, the discriminant is always positive. On the contrary, if $|\omega'| \le \sqrt{3}$, we see that reducing $R'$ amounts to reducing $q$ and may lead to a negative discriminant.
The natural frequencies are centered around $\omega=0$, so the normalized frequencies $\omega'$ are centered around $-\eta/b$.
% which is close to 0 if $k$ is small. 
If the magnitude of this value is larger than $\omega'_{\rm c}=\sqrt{3}$, the majority of oscillators have a positive discriminant. 
This leads to a first very crude criterion for a synchronized state composed of a majority of synchronized oscillators:
\begin{equation}
k \ge \hat{k}_{\rm c} \equiv \frac{\mu-\sqrt{3}\Omega}{2\cos(\alpha-\pi/3)}\,,
\label{eq:kc2:simple}
\end{equation}
which does not take into account the width $\sigma$ of the natural frequencies distribution, neither its tails.
To get a better estimate, we compute the fraction $P$ of oscillators which have $|\omega'| \ge \omega'_c$:
$$
P = 1 - \int_{\eta-b\omega'_{\rm c}}^{\eta+b\omega'_{\rm c}} g(\omega) {\rm d}\omega \,.
$$
For a Lorentzian distribution, this fraction can be rewritten as:
$$
\cotan(\pi P) = \frac{(b\omega'_{\rm c})^2 - \sigma^2 - \eta^2}{2b\sigma\omega'_{\rm c}} \,.
$$

An heuristic condition for the existence of a partially synchronized macrostate containing a majority of synchronized oscillators is obtained by requiring $P\ge 1/2$. This implies $\cotan(\pi P) \le 0$ which is ensured if $\sigma^2 + \eta^2 \ge (b\omega'_c)^2$.
We then rewrite this condition assuming $\Omega=0$ for the sake of simplicity. This approximation is supported by the DNS where we observe that $\Omega$ is always much smaller than $k\sin\alpha$, and so $\eta \simeq k\sin\alpha$.
Replacing $\eta$ and $b$ by their linear expression in $k$, we obtain
$k\ge \hat{k}_{1/2}$ with
\begin{equation}
\hat{k}_{1/2} \cos \alpha  = \frac{\mu - \sqrt{3}\sqrt{\sigma^2+ (\mu^2-3\sigma^2)\tan^2 \alpha}}
{1-3\tan^2\alpha}
\label{eq:kc2}
\end{equation}
The specific values $\hat{k}_{1/2}$ or $\hat{k}_{\rm c}$ are not estimates of $k_{\rm sync}$, but simple heuristic bounds for the existence of a particular mixed macrostate composed of a majority of synchronized oscillators, based on DNS observations and conjectures. 
These bounds are independent of the system size. As expected ---~and shown below in Fig.~\ref{fig:phase.plane}~--- they are both lower than the estimate $\hat{k}_{\rm sync}$ obtained by solving the auto-coherence relation for the fully synchronized state.
Decreasing the width $\sigma$ of the frequency distribution increases  $\hat{k}_{1/2}$ up to $\hat{k}_{\rm c}$. 
Requiring a fraction $P$ larger than 1/2 increases $\hat{k}_{1/2}$ which can make it larger than $\hat{k}_{\rm inc}$ for small $\alpha$.
%

%The discriminant of (\ref{eq:cubic:rho}) is negative for $R'<\sqrt{2}$ in a compact interval of  $\omega'^2$ and two additional real solutions exist which have a much smaller amplitude ($\rho<2$) than the former single solution.
%In that situation, there are two bands of oscillators in the frequency distribution which can have an amplitude $\rho<3$ and can therefore reduce the mean field amplitude $R'$ according to (\ref{eq:autocoherence:real}).
%
%The fraction $D$ of such oscillators amongst the complete distribution is bounded from above by the fraction of oscillators with $p\le0$ in eq.(\ref{eq:cubic:rho}), {\em i.e.} $|\omega'|\le\sqrt{3}$.

%{For values of the coupling coefficient {$k$ larger than $\hat{k}_{\rm c}$ or $\hat{k}_{1/2}$}, we are sure that macrostates with $R'>\sqrt{2}$ are composed of a majority of oscillators with $\rho>3$ {which contribute} positively to the mean field according to~(\ref{eq:autocoherence:real}). On the contrary, below $\hat{k}_{1/2}$, the situation is unclear and we cannot characterize the macrostate.}

This counting of oscillators (requiring 50\% of the oscillators to have $p>0$)  is very crude as it does not take into account the order parameters $R'$, which may change the sign of the discriminant for oscillators with $p<0$. These oscillators are the most numerous when $\alpha$ is small because the center of the frequency distribution $\{\omega'\}$ is then close to 0.
Nevertheless, for large $\alpha$, it shows heuristically that a synchronized state can exist where the completely incoherent macrostate is linearly stable, which is in strong contrast to the Kuramoto phase model.
We haven't achieved yet the stability analysis of this synchronized macrostate and cannot explain why it is not dynamically selected by the system for smaller $k$ and $\alpha$ values.

\section*{Discussion}

\begin{figure}[tb]
\begin{center}
\includegraphics[width=\linewidth]{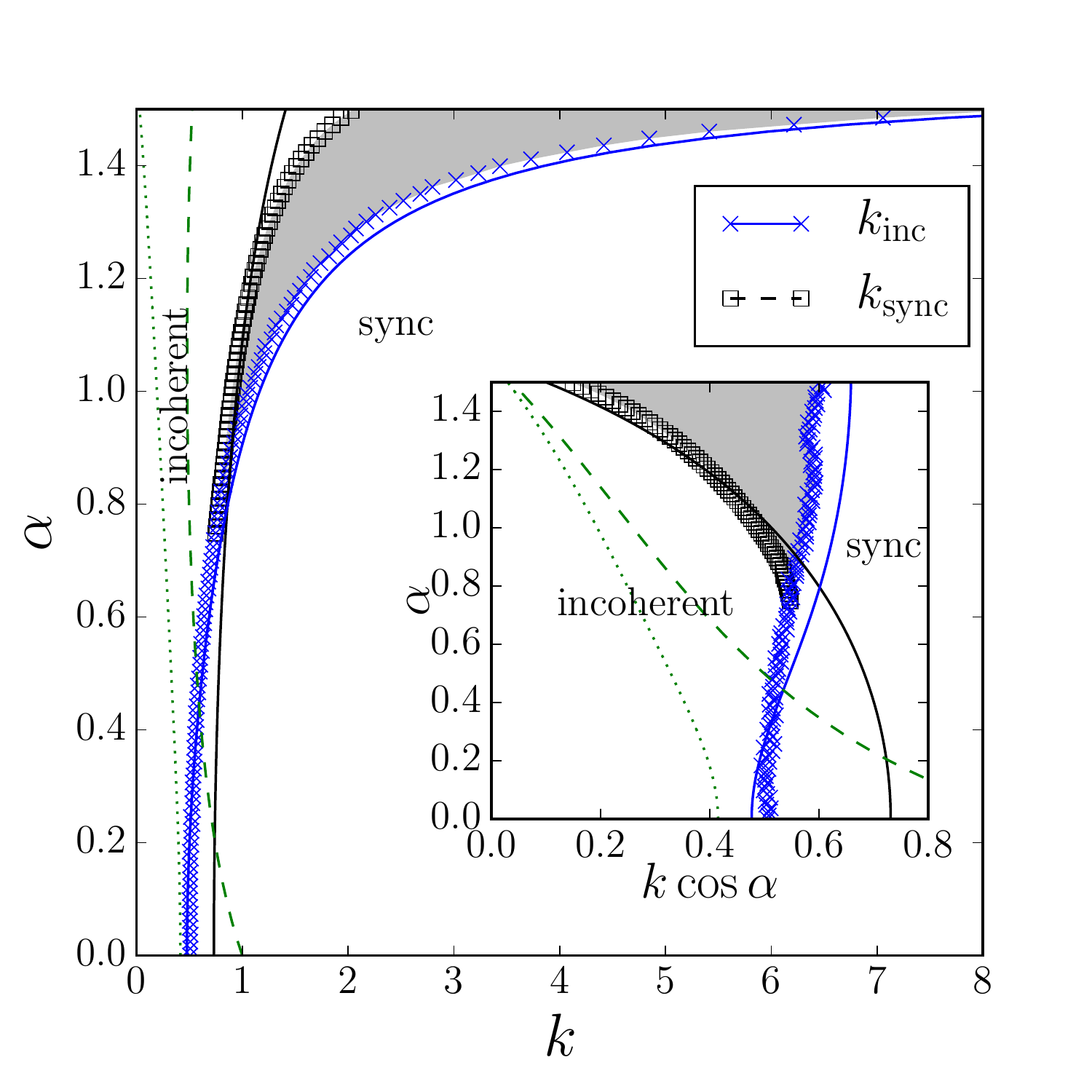}
\end{center}
\caption{Phase diagram.
Blue crosses, resp. black squares, represent $k_{\rm inc}$, resp. $k_{\rm sync}$.
Solid blue line is $\hat{k}_{\rm inc}$ from eq.(\ref{eq:kc1}).
Solid black line is $\hat{k}_{\rm sync}$ from solving eqs.(\ref{eq:autocoherence:real},\ref{eq:autocoherence:imag}). 
Green dashed line is $\hat{k}_{c}$ from eq.(\ref{eq:kc2:simple}) with $\Omega=0$.
Green dotted line is $\hat{k}_{1/2}$ from eq.(\ref{eq:kc2}).
The gray region indicates the hysteresis observed in the DNS. 
The inset uses a different representation to give a better picture of the small $\alpha$ region.
}
\label{fig:phase.plane}
\end{figure}

Increasing $\alpha$ makes the synchronization transition discontinuous. In Fig.~\ref{fig:phase.plane}, we present a phase diagram in the $(\alpha, k)$ plane, obtained by DNS with a fixed realization of $\{\omega_j\}$. \N{At the} special value $\alpha_{\rm c}=0.75$ where $k_{\rm inc}=k_{\rm sync}$ \N{one expects a codimension-two bifurcation~\cite{Crawford:1999} which is equivalent to} a critical point in phase transitions. Above $\alpha_{\rm c}$, the width of the hysteresis increases with $\alpha$. 
Below $\alpha_{\rm c}$ one cannot easily measure $k_{\rm sync}$ from DNS, unless performing a detailed examination of the time fluctuations of $R$ \N{or of the composition of the microstates}.
 
For coupling values $k$ slightly larger than $\hat{k}_{\rm inc}$, the incoherent macrostate is linearly unstable, so the system has to be \N{in} a synchronized macrostate. The fraction of synchronized oscillators in the corresponding microstates is unknown.
On the contrary, ensuring the linear stability of the incoherent macrostate is not a sufficient criterion for discarding the existence of a synchronized macrostate~\cite{Strogatz:2000} which may turn out to be stable. This is exactly the situation we observe for $\alpha>\alpha_{\rm c}$.
 
We can therefore interpret the discontinuous transition as resulting from the existence of 
either a fully or partially synchronized state containing a non-vanishing number of synchronized oscillators for coupling values where the completely incoherent macrostate is stable.
In fact, the depiction of the phase diagram in the inset of Fig.\ref{fig:phase.plane} using the real part $k\cos\alpha$ of the coupling instead of its magnitude $k$ suggests that the hysteresis region results from the incursion of a particular synchronized state in the incoherent region (defined with formula (\ref{eq:kc1})). 
We have not studied the stability of the specific macrostates we have considered (whether fully or partially synchronized). Our extensive DNS explorations shown that for smaller values of $\alpha$ and weak coupling, the system selects the completely incoherent solution, and the classical Kuramoto states ---~which have a vanishing number of synchronized oscillators around $k_{\rm inc}$~--- thus giving a continuous transition to synchronization. For larger values of $\alpha$, where we report bistability, we never observe classical Kuramoto states and the fraction of synchronized oscillators is always larger than 50\%.

\begin{figure}[tb]
\begin{center}
\includegraphics[width=\linewidth]{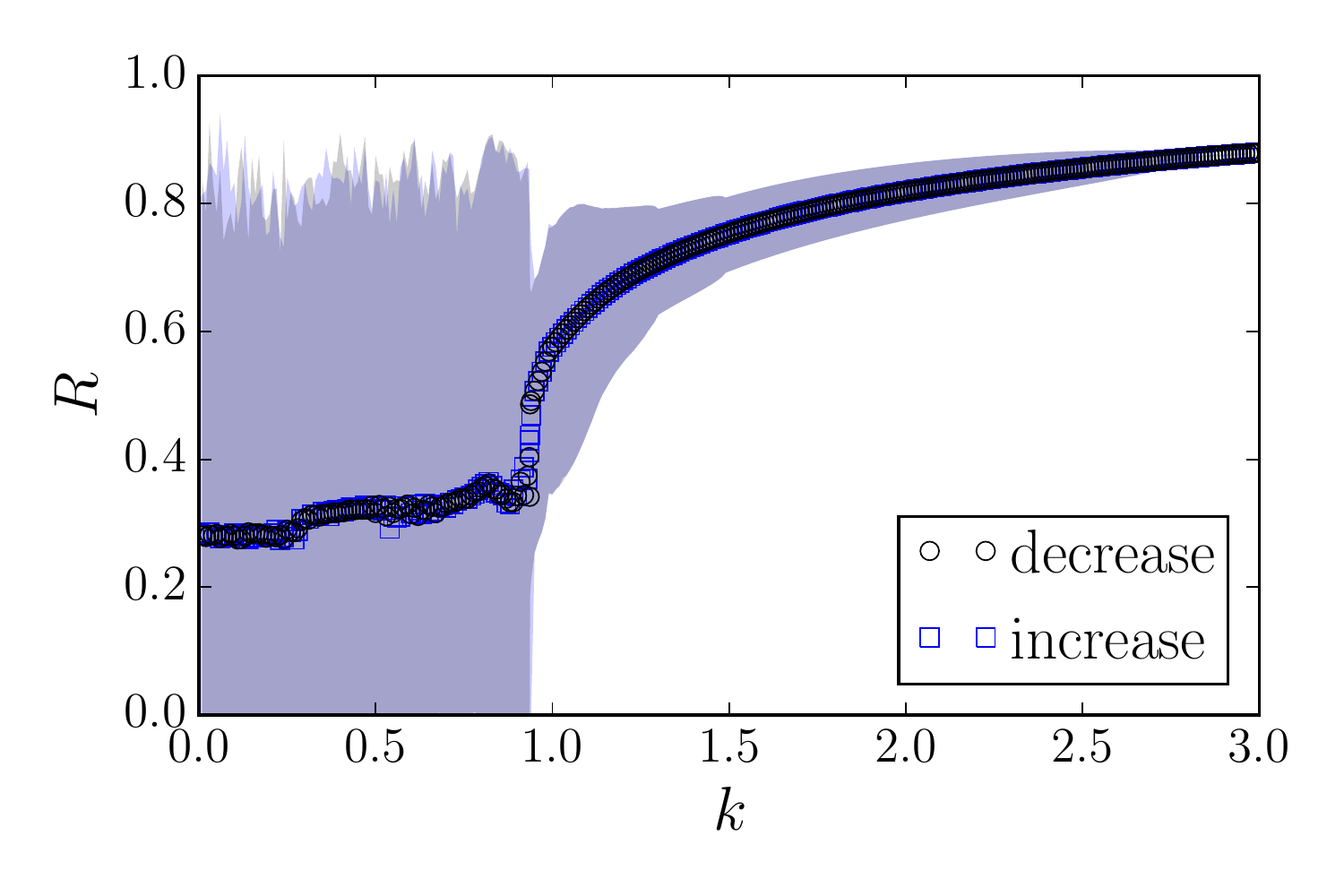}
\end{center}
\caption{Bifurcation diagram for $\alpha=1.3$, in a small system composed of $N=10$ oscillators.
Symbols indicates DNS values of the time average of ${R}$, while the range of time fluctuations is represented by color shades. Blue squares, resp. black circles, correspond to macrostates obtained by adiabatically increasing, resp. decreasing, the coupling $k$.
The transition is continuous.
}
\label{fig:small:system}
\end{figure}

Fig. \ref{fig.size_effect} shows that reducing the system size \N{$N$}, which reduces the dimensionality of the problem, impacts the stability of the incoherent state but not the existence of the fully synchronized state. 
\N{As $N$ is reduced, the expected $k_{\rm inc}$ ---~defined as the average of $k_{\rm inc}$ over many realizations of the disorder $\{\omega_j\}$~--- is reduced whereas $k_{\rm sync}$ is unchanged. 
This suggests that although the transition is discontinuous in the thermodynamic limit, for small enough systems $N \lesssim 18$ (obtained by extrapolating the linear fit in Fig.~\ref{fig.size_effect}), the value $k_{\rm inc}$ can be lower than $k_{\rm sync}$
As a consequence,} small systems, {\em e.g.} a few dozens oscillators or clocks, probably do not exhibit a discontinuous transition to synchronization.
\N{To check this, we performed some DNS for $N=10$, $\alpha=1.3$ and indeed observed a continuous transition, 
\Q{as can be seen in Fig.~\ref{fig:small:system}}.
Increasing or decreasing the coupling does not change the observed macrostate: there is no more hysteresis.
The transition occurs for a value of the coupling $k \simeq 0.9$ which is smaller than $k_{\rm sync}\simeq 1.25$. 
Not only time fluctuations of the order parameter $R(t)$ are large when $N$ is small 
\Q{(shaded regions in Fig.~\ref{fig:small:system})} 
but also fluctuations of $k_{\rm inc}$ from one replica to another (see the increasing error bars in Fig.~\ref{fig.size_effect}, defined as the standard deviation).
So we may except that even when $N$ is small, there exist some very specific replicas where $k_{\rm inc}$ is larger that $k_{\rm sync}$ so that the transition is discontinuous.}

Other discontinuous transitions to synchronization have been reported recently. In the Kuramoto-Sakaguchi phase system~\cite{Wolfrum:2013} \N{---~a generalization of the Kuramoto model where an angular shift $\alpha$ is introduced in the coupling term, exactly as in our eq.(\ref{eq:polar:phase}), and acts as a frustration parameter~---}
a detailed mathematical analysis has shown that a bimodal frequency distribution is not required, and an unimodal frequency distribution can be specifically designed to produce bi-stability. In that case, the phase of the oscillators is forced to be aligned to $\Psi+\alpha$ by virtue of eq.(\ref{eq:polar:phase}) whereas the mean field builds itself at $\Psi$ by definition (\ref{eq:mean_field}). When the tails of the frequency distribution are populated enough, these two apparently contradictory dynamics can equilibrate in a mixed microstate with a non-vanishing number of synchronized oscillators, even if the incoherent state is still stable. This reasoning applies directly to our case when considering that the angle-frequency relation (\ref{eq:polar:phase}), analogous to the Kuramoto-Sakaguchi phase equation, is perturbed by the extra amplitude variable $r$ via eq.(\ref{eq:polar:radius}) which modifies the weights of frequencies, and hence the effective frequency distribution.

\N{The Kuramoto model with inertia (KI)~\cite{Tanaka:97,Acebron:2000,Olmi:2014} is another paradigm where hysteretic transitions to synchronization have been reported in detail and carefully analyzed; recent studies have even taken into account a frustration in the coupling~\cite{Barre:2016} or the presence of noise~\cite{Gupta:2014}. 
Taking into account a finite inertia amounts to considering the frequency $\dot{\theta}$ as an additional variable~\cite{Acebron:2000}, which doubles the number of freedom degrees, as does adding the radius variable in the Stuart-Landau (SL) model. 
Nevertheless, observed regimes are different.
In our case, we found an hysteretic transition between the completely incoherent state and a single, unique, synchronized macrostate. In the Kuramoto model with inertia, the situation is more complex as there are additional hysteretic transitions between multiple synchronized macrostates.
At the microscopic level, oscillators dynamical equations have fixed points or limit cycles.
%
%BELOW IS WRONG! TO EDIT
In the KI model, two stationary fixed points (one stable and one unstable) for locked oscillators can co-exist with a limit cycle for incoherent oscillators (see eq.(6) in ref\cite{Tanaka:97} or eq.(3) in ref\cite{Olmi:2014}) depending on initial conditions. In the SL model, solving the cubic equation (\ref{eq:polar:radius}) for the radius may give up to three solutions, while the limit cycle corresponding to an incoherent state is difficult to express. 
Nevertheless, for $R'>\sqrt{2}$, the fixed point is unique and the limit cycle doesn't exist; and we have not observed in the DNS any sensitivity of the solution (fixed point or limit cycle) on the initial conditions, but rather on $R'$ and on the natural frequency.
So, although both KI and SL model may lead to discontinuous transitions to synchronization, the situation appears simpler in the Stuart-Landau model (at least for $c_2=0$), as it originates at the global level (when solving the auto-coherence relation) only, whereas it is already present at the microscopic level in the Kuramoto model with inertia.
}

Beyond mean-field coupling, the topology of the underlying network (totally absent in the classical Kuramoto formulation as well as in our case) can also allow a specific synchronized state to exist for values of the coupling where the incoherent macrostate is still stable. This may explain the discontinuous transition ---~also labeled "explosive"~--- reported in such systems~\cite{Gomez:2011} when, e.g., the natural frequency of a node is made dependent on its degree. This modifies strongly the distribution of effective frequencies, in the same ways as the amplitude of oscillators does in our case. If the dynamical equations for a single oscillator can be manipulated to obtain a criterium for the synchronization of this oscillator, or if an heuristic criterium is available, it is then in theory possible to estimate the fraction of oscillators in a given distribution that satisfies the criterium and hence to deduce some bounds for the existence of synchronized state with a finite, non vanishing, order parameter.

Because of the plurality of possible macrostates~\cite{Olmi:2014}, and hence the possible "high" multistability of the system, we prefer to qualify the transition as continuous or discontinuous rather than using the second or first order phase transitions analogy, although they may be proven one day to be correct.

\section*{Conclusion}

We studied the transition to synchronization in a model system composed of globally coupled Stuart Landau oscillators, for which not only the phase, but also the amplitude is considered. When the imaginary part of the coupling ---~or equivalently the ratio between reactive and diffusive coupling~--- is increased, the transition becomes discontinuous. To understand this new feature, we considered specific macrostates of the system and showed that they can coexist in some range of parameters.

We studied the stability of the completely incoherent macrostate, as well as the existence of a synchronized macrostate composed either of a totality of a majority of synchronized oscillators. We argued that the existence of such a synchronized state allows a discontinuous transition when the incoherent state becomes unstable. A stability analysis of these synchronized macrostates is still missing and is under progress, albeit difficult. 
The existence of such specific synchronized macrostates is directly related to the distribution of the effective frequencies. In our case, the natural frequency distribution is modified ---~with respect to the Kuramoto model where only the phase dynamics is prescribed --- by the additional amplitude variable for each oscillator: this extra freedom degree not only shifts the distribution, but also gives more weight to oscillators within a given frequency band. 
We also numerically studied the finite size effects and suggested that reducing the system size can result in the disappearance of the discontinuous transition. 
We discussed recent results in similar systems and argued that our description is generic.

%A detailed study of all possible microstates is under progress.

\acknowledgments

Authors thank Zonghua Liu for interesting discussions. This work benefited from a JoRISS grant from ENS de Lyon and East China Normal University, and a CMIRA Accueil'Sup grant from R\'egion Rh\^one-Alpes.

\end{document}